\documentclass[submit]{pepsart}

\usepackage{amsthm,amsmath}
\usepackage[utf8]{inputenc} 
\usepackage{wasysym}
\usepackage[round,authoryear]{natbib}
\usepackage{longtable}
\usepackage{footnotehyper}
\makesavenoteenv{longtable}
\usepackage{chemfig}
\usepackage{graphicx}
\usepackage{color}
\usepackage{tablefootnote}

\startlocaldefs
\def\cite{\citep}
\endlocaldefs

\begin{document}

\begin{frontmatter}

\begin{fmbox}
\dochead{Research}


\title{Characteristics of temporal variability of long-duration bursts of high-energy radiation associated with thunderclouds on the Tibetan plateau}


\author[
   addressref={aff1},                   
   corref={aff1},                       
   email={tsuchiya.harufumi@jaea.go.jp}                
]{\inits{H.}\fnm{Harufumi} \snm{Tsuchiya}}
\author[
   addressref={aff2},
   email={hibino@n.kanagawa-u.ac.jp}
]{\inits{K.}\fnm{Kinya} \snm{Hibino}}
\author[
   addressref={aff3},
   email={kawata@icrr.u-tokyo.ac.jp}
]{\inits{K.}\fnm{Kazumasa} \snm{Kawata}}
\author[
   addressref={aff3},
   email={ohnishi@icrr.u-tokyo.ac.jp}
]{\inits{M.}\fnm{Munehiro} \snm{Ohnishi}}
\author[
   addressref={aff3},
   email={takitai@icrr.u-tokyo.ac.jp}
]{\inits{M.}\fnm{Masato} \snm{Takita}}
\author[
   addressref={aff4},
   email={kmuna00@shinshu-u.ac.jp}
]{\inits{K.}\fnm{Kazuoki} \snm{Munakata}}
\author[
   addressref={aff4},
   email={ckato@shinshu-u.ac.jp}
]{\inits{C.}\fnm{Chihiro} \snm{Kato}}
\author[
   addressref={aff5},
   email={sshimoda@riken.jp}
]{\inits{S.}\fnm{Susumu} \snm{Shimoda}}
\author[
   addressref={aff6},
   email={sqq@sdu.edu.cn}
]{\inits{Q.Q.}\fnm{Quanqi} \snm{Shi}}
\author[
   addressref={aff6},
   email={wangshuo_wh@sdu.edu.cn}
]{\inits{S.}\fnm{Shuo} \snm{Wang}}
\author[
   addressref={aff6},
   email={chenyao.han@mail.sdu.edu.cn}
]{\inits{C.Y.}\fnm{Chenyao} \snm{Han}}
\author[
   addressref={aff7},
   email={zhailiuming@nao.cas.cn}
]{\inits{L.M.}\fnm{Liuming} \snm{Zhai}}



\address[id=aff1]{
  \orgname{Nuclear Science and Engineering Center, Japan Atomic Energy Agency}, 
  \street{2-4, Shirakata, Tokai-mura},                    %
  \city{Naka-gun, Ibaraki}                          
  \postcode{319-1195},                           
  \cny{Japan}                                    
}
\address[id=aff2]{%
  \orgname{Faculty of Engineering, Kanagawa University},
  \city{Yokohama, Kanagawa}
  \postcode{221-8686},
  \cny{Japan}
}
\address[id=aff3]{%
  \orgname{Institute for Cosmic Ray Research, University of the Tokyo},
  \street{5-1-5, Kashiwanoha},
  \city{Kashiwa, Chiba}
  \postcode{277-8582},
  \cny{Japan}
}
\address[id=aff4]{%
  \orgname{Department of Physics, Shinshu University},
  \city{Matsumoto, Nagano}
  \postcode{390-8621},
  \cny{Japan}
}

\address[id=aff5]{%
  \orgname{RIKEN},
  \street{2-1, Hirosawa},
  \city{Wako, Saitama}
  \postcode{351-0198},
  \cny{Japan}
  }
  \address[id=aff6]{%
  \orgname{Shandong Provincial Key Laboratory of Optical Astronomy and Solar-Terrestrial Environment, Institute of Space Sciences, School of Space Science and Physics, Shandong University},
  \city{Weihai}
  \postcode{264209},
  \cny{China}
  }
  \address[id=aff7]{%
  \orgname{National Astronomical Observatories, Chinese Academy of Sciences},
  \city{Beijing}
  \postcode{100101},
  \cny{China}
  }


\begin{artnotes}
\note{Characteristics of temporal variability of long-duration bursts}     
\note[id=n1]{Equal contributor} 

\end{artnotes}

\end{fmbox}


\begin{abstractbox}

\begin{abstract} 
From 1998 to 2017, neutron monitors located at an altitude of 4300 m on the Tibetan plateau detected 127 long-duration bursts of high-energy radiation in association with thunderclouds. These bursts typically lasted for 10 to 40 minutes, and 89\% of them occurred between 10:00 and 24:00 local time.
They were also found to be more likely to occur at night, especially during 18:00$-$06:00 local time period. The observed diurnal and seasonal variations in burst frequency were consistent with the frequencies of lightning and precipitation on the Tibetan plateau. 
Based on 19 years of data, the present study suggests that
an annual variation in burst frequency has a periodicity of $\sim$16 years and a lag of $\sim$3 years relative to solar activity.

\end{abstract}




\end{abstractbox}
%

\end{frontmatter}

\section{Introduction}
In the 1920s, C.~T~.R. Wilson proposed that electrons in thunderclouds are accelerated to high energy in strong electric fields and produce x-rays by bremsstrahlung~\citep{wilson_1924,wilson_1925}. 
Since then, 
many observations have shown that lightning and thunderclouds generate
such x-ray signals, confirming that they are indeed powerful particle accelerators ~\citep[][and references therein]{dwyer_high-energy_2012}. 
Beyond the Wilson prediction, bremsstrahlung gamma-rays with a few tens of MeV
have been detected from lightning and thunderclouds. The detection of the high-energy photons
suggests that lightning and thunderclouds can accelerate electrons beyond a few tens of MeV
inside them. Theoretical works~\cite[{e.g.}][]{gurevich_runaway_1992, dwyer_fundamental_2003,dwyer_relativistic_2007,babich_fundamental_2004,babich_source_2010} have
pointed out that the production of those high-energy radiations is due to relativistic runaway electrons. 
Seed electrons produced by, for example, cosmic rays are thought to be accelerated to relativistic energies
when electric fields exceed a critical threshold and become such runaway electrons. 

Solar activity influences the heliospheric magnetic field, resulting 
in variations in the intensity of galactic cosmic rays reaching the 
Earth in an approximately 11-year cycle. Therefore, it is thought that 
solar activity, by affecting the flux of galactic cosmic rays reaching 
the Earth, also impacts the production of runaway electrons.
However, it remains unclear whether cosmic rays or solar activity play a role in producing the
runaway electrons responsible for high-energy bremsstrahlung gamma-rays. 

Bursts of high-energy radiation during thunderstorms can be categorized into two types when classified simply by their duration.
One type is characterized by intense short-duration bursts lasting for microseconds to milliseconds. 
The other type is a long-duration burst, which lasts from a few seconds to a few tens of minutes and exhibits relatively fainter 
characteristics compared to the short-duration bursts. 
Numerous measurements of these short-duration bursts have been made by space- and ground-based detectors and revealed several characteristics including the source altitude and their relation to lightning.
In contrast to short-duration bursts, 
long-duration bursts such as gamma-ray glows or terrestrial ground enhancements (TGEs)
have been observed less frequently until recently. 
However, observations made in the coastal area of the Japan Sea~\cite{torii_observation_2002,tsuchiya_detection_2007,tsuchiya_long-duration_2011,wada_meteorological_2021} at high mountains~\cite{torii_gradual_2009,tsuchiya_observation_2009,chilingarian_groundbased_2010,chilingarian_particle_2011,tsuchiya_observation_2012} and at high aircraft altitude~\cite{ostgaard_glow_20km_2019,kochkin_rapid_2021} have led to increased detection of these long-duration bursts and have drawn attention to their characteristics, including their duration and their correlation with meteorological phenomena.

The increasing amount of data on long-duration bursts have allowed us to compile a comprehensive catalog of such phenomena and to gain insight into their characteristics at various observation sites, including high mountain regions~\cite{chilingarian_catalog_2019} 
and coastal areas of the Japan Sea~\cite{wada_catalog_2021}.
However, more data are needed to explore whether there is a correlation between long-duration bursts and cosmic rays and to understand the meteorological and geographical factors that contribute to the generation of these bursts. 
By studying these conditions, we can identify commonalities among long-duration bursts and develop a deeper understanding of their production mechanisms.
To this end, we will present observational results
obtained by neutron monitors (NM) at Yangbajing on the Tibetan plateau (TP) during 1998$-$2017, and then discuss the temporal variability of long-duration bursts. 

\section{Experiments}
\subsection{Overview}
The Yangbajing NM has been in operation at the Yangbajing cosmic-ray observatory for more than two decades. 
The observatory is located at an altitude of 4300 m (corresponding to an atmospheric depth of $\sim$600 $\mathrm{gcm^{-2}}$) in the TP. 
In addition, electric field mills (EFMs) have been deployed at the observatory in 2010
to study possible correlations between variations in count rate measured by a detector such as the Yangbajing NM and electric fields (EFs)~\cite{amenomori_observation_2013}.  
To clarify the points of the present work, we briefly explain their characteristics below.

\subsection{Neutron monitor}
The main purpose of the Yangbajing NM is to investigate the ion acceleration at the solar surface and to monitor galactic cosmic-ray intensity associated with the solar activity ~\cite{miyasakaSolarEvent202005,tsuchiya_upper_2007}. In the present study,
the NM is repurposed to examine the production mechanism of long-duration bursts of high-energy radiation associated with thunderclouds. 
Hereafter, the term "long-duration burst" is employed to refer to the high-energy radiation detected by the Yangbajing NM in association with thunderclouds.

As shown in Figure~\ref{fig:NMscmatic},  
the Yangbajing NM consists of 28 NM-type $\mathrm{BF_3}$ counters surrounded by lead blocks and polyethylene in the standard NM64 configuration.
Each NM unit has one $\mathrm{BF_3}$ counter with dimensions of 190 cm in length and 15 cm in diameter. The total area of the Yangbajing NM is 32 $\mathrm{m^{2}}$.
The outermost polyethylene serves the purpose of moderating incoming fast neutrons, while
the embedded lead blocks breed the moderated neutrons via (n, $x$n) reactions to enhance
the detection efficiency for neutrons ($x$ represents a number). 
When entering the $\mathrm{BF_3}$ counters, the moderated neutrons undergo the (n, $\alpha$) reaction:
$\mathrm{n} + \mathrm{^{10}B} \rightarrow \alpha + \mathrm{^{7}Li}$.
The cross sections of this reaction become larger as the neutron energy approaches the thermal energy, thus allowing efficient detection of neutrons in thermal to epithermal energy ranges. 
To further improve neutron detection efficiency, 
the isotopic composition of $^{10}\mathrm{B}$ in each $\mathrm{BF_3}$ counter is enriched 
to 98\% from a natural isotopic ratio of $\sim$20\%.
Charged particles produced by the above reactions lose energy due to the ionization loss in gas, resulting
in anode signals in the counter. A data acquisition system collects the output signals from each counter and registers the number of signals every one second.

In our previous study~\cite{tsuchiya_observation_2012}, 
we determined the proportions of individual particles that contribute to the total count rate of the Yangbajing NM.  
Secondary cosmic-ray neutrons and protons dominate the total count rate with fractions
of 83\% and 12\%, respectively. In contrast, leptons, including gamma rays and electrons, 
contribute to it with only 5\% [see also \citet{clem_dorman_nm_2000}].
The previous study also indicated that the same type of NMs as the Yangbajing NM
have a small but non-negligible detection efficiency for gamma rays through photonuclear reactions with lead
blocks ($\gamma + ^\mathrm{A}\mathrm{Pb} \rightarrow \mathrm{n} + ^\mathrm{A-1}\mathrm{Pb}$) (A$=$206,207,208), with a threshold
gamma-ray energy of 7 MeV~(IAEA, 2000). 
Given these insights, we demonstrated that  bremsstrahlung photons
were responsible for the observed count increases during thunderstorms in the Yangbajing NM, rather than neutrons produced by 
photonuclear reactions in the atmosphere [e.g. \citet{enoto_photonuclear_2017}]. 
Therefore, for the present study as well, we assume that count increases of the Yangbajing NM can be attributed primarily to bremsstrahlung photons with energy higher than 7 MeV. 

\subsection{Electric field mill}
Electric field data play a crucial role in understanding the association of observed increases with thunderclouds and lightning.
For this purpose,
two BOLTECK EFM-100 commercial electric field mills have been operational at the observatory since 2010,
positioned 25 m apart. One is situated at ground level, while the other
is mounted atop an experimental building. The vertical separation between them is
approximately 3.4 m. The signals from both EFMs are transmitted to computers within the building via optical cables. 
These signals are recorded at an interval of every 0.1 s, providing field strength below 
100 $\mathrm{kVm^{-1}}$, with a measurement resolution of 20 $\mathrm{Vm^{-1}}$.

\section{Analysis}
To identify long-duration bursts, we performed several steps of analysis of data collected by the Yangbajing NM from 1998 to 2017. 
First, we made a 5-minute count history corrected for atmospheric pressure variations for each day. Pressure
fluctuations typically induce count changes within $\pm2$\%. Second, we constructed an occurrence histogram of 
NM counts in every 5-minute count history, $N_\mathrm{5min}$, and fitted a Gaussian function to this histogram to determine the mean value $N_\mathrm{BG}$ and the standard deviation $\sigma_\mathrm{1day}$ of the Gaussian function
every day. 
The resulting $\sigma_\mathrm{1day}$ served as an estimate of the uncertainty in the 5-minute counts for each day. 
Third, we calculated the statistical significance, $\sigma_\mathrm{s}$, for each $N_\mathrm{5min}$ using the formula;
\begin{equation}
\sigma_\mathrm{s} = \frac{N_\mathrm{5min} - N_\mathrm{BG}} {\sqrt{N_\mathrm{5min} + \sigma_{\mathrm{1day}}^2}}.
\end{equation}
This allowed us to assess whether the daily NM data contained count bins with $\sigma_\mathrm{s}> 4$. 
The probability of encountering a bin with a significance of 4 or higher by chance is $5\times10^{-5}$. Therefore, 
out of the 288 bins in each 5-minute count history, 
the number of bins with a significance of 4$\sigma$ or greater is only 0.014. 
Among the $N_\mathrm{5min}$ counts, we defined the most prominent one as the peak count, $N_\mathrm{p}$, in each 5-minute count history. This data set, which we refer to as the "potential burst set", contains 135 events that represent possible long-duration bursts associated with thunderclouds. 

From the potential data set, we extracted a 5-minute count history over a time span of $\pm$2h
centered on the peak time "$t_\mathrm{p}$" when $N_\mathrm{p}$ was recorded. This time interval of $\pm$2h is
sufficient to cover the entire duration of each burst and to estimate the background level of
each burst. 
To define the burst duration "$\Delta t_\mathrm{d}$", we identified a continuous time range in which the statistical significance $\sigma_\mathrm{s}$ remained above 1 before and after the peak time $t_\mathrm{p}$. 
By excluding all count bins in this $\Delta t_\mathrm{d}$, 
two-time regions are left in every extracted count history to estimate the background. 
The background level was determined by fitting count bins in the two time regions
with a quadratic function. The background level in $\Delta t_\mathrm{d}$ was then
estimated by interpolating the derived quadratic function. 

After completing the above analysis, an eye scan was performed to detect any suspicious events that may have been caused by irregular factors such as power failures. As a result of this scan, eight events were identified as suspicious events in the potential burst set. Subsequently, a total of 127 long-duration bursts were selected from the NM data collected between 1998 and 2017. This data set is referred to as the final burst set.  None of the events in the final burst set was found to be associated with solar energetic particle events, including ground-level enhancements~\cite{ramiGLE2017,cohenGLE2018}, suggesting that such a solar particle event does
not contribute to the production of long-duration bursts.

The current analysis is limited in the detection of faint bursts, because it only considers those with $>$4$\sigma$. 
This selection criterion is one of the reasons for the detection of only 127 events during the long observation period (1998$-$2017). This frequency is also likely influenced by the higher NM threshold of 7 MeV and the limited spatial illumination of a long-duration burst. Gamma-ray detectors using NaI and BGO scintillators can readily capture long-duration bursts with energies of a few MeV or less~\cite{chilingarian_catalog_2019,wada_catalog_2021}, while 
these low-energy gamma rays are not detectable by the Yangbajing NM. In addition, it is known that a long-duration burst illuminates only a limited area ($0.5-1$ km) on the ground~\cite{tsuchiya_long-duration_2011, wada_gamma-ray_2019,Tsurumi_GRL_2023}. 
At the altitude of 4300 m, the mean free path of gamma rays with
energy of 10$-$50 MeV is approximately 700$-$900 m.
Therefore, it would be more difficult to clearly detect long-duration bursts originating significantly farther away from NMs than the mean free path.

\section{Results and Discussion}
\subsection{Example of time profiles of long-duration bursts}
Figure~\ref{fig:exampleLCs} depicts examples of the 5-minute count histories in the final burst set. It is evident that the individual count histories exhibit significant count increases.
Table~\ref{tab:eventslist} presents several key characteristics for each of 127 long-duration bursts, 
including its net count increase denoted as "$N_\mathrm{tot}$" and the related statistical significance. The $N_\mathrm{tot}$
is calculated by subtracting the interpolated background from the total observed count in $\Delta t_\mathrm{d}$.
The observed count increases are superimposed on gradual count changes caused primarily by solar diurnal variation, 
which typically ranges within $\pm0.3$\% in amplitude. 
This variation arises from the 
anisotropy of galactic cosmic rays below several tens of GeV in interplanetary space.

Unlike the gradual changes induced by the solar diurnal variation, EF variations result in faster changes in the NM count rate. Figure~\ref{fig:TwoCntWithEFMex1} shows examples of the 
relationship between EF changes and observed count increases.
The EF intensity remains at $\sim$0.1 $\mathrm{kVm^{-1}}$  
under fair weather conditions, with its sign following atmospheric physics conventions.
A positive value denotes an EF directed downward from the sky to the ground, and vice versa.
The EF is found to exhibit significant variations lasting for 1$-$1.5h. While the count increases in the events
(Figure~\ref{fig:TwoCntWithEFMex1}) seem to coincide with periods of relatively large positive EFs, the peak times of these count increases do not necessarily align with the peak times of the positive EFs. A possible explanation could be that the alterations in the EF within a thundercloud, responsible for electron acceleration, may not precisely synchronize with the observed variations in the EF at the ground surface. Intense, sharp peaks are also found in EF variations (e.g., 20150827 and 20160617) and
indicate the presence of lightning. These lightning events causing significant peaks in EF variations do not appear to impact the observed long-duration bursts. This is probably because the location of the lightning observed by the EFM is far from NMs. Therefore, when investigating the relationship between EF and long-duration bursts, it would be necessary to specify not only the EF intensity but also the distance to a thundercloud or lightning. This will be our future task.

All other events\footnote{Comparisons between count increases and EF variations are available in supplement information.} detected after the EFM installations
are also found to have occurred under similar EF conditions, suggesting a common underlying mechanism to produce a long-duration burst.
Typically, a thundercloud is considered to have a tripole charge structure, comprising a lower positive charge region (LPCR), a middle negative charge region, and an upper positive charge region~\cite{Cooray_book}. 
Electrons, located between the LPCR and the middle negative charge region, undergo the electric-filed acceleration toward the LPCR and emit high-energy photons through bremsstrahlung directed to the ground.
Based on measurements carried out in the TP at an altitude of 4500 m above sea level (a.s.l.),
it was found that the cloud base altitude is generally 1 km
above the ground surface~\cite{qie_lower_2005}. 
In our previous study~\cite{tsuchiya_observation_2012}, we examined a long-duration burst that occurred on 22 July 2010 (listed as 20100722 in Table~\ref{tab:eventslist}), 
and found that about 90\% of the NM count enhancement was 
caused by bremsstrahlung photons emitted by electrons accelerated in thunderclouds.  
This estimation assumed a cloud base altitude of typically about 1 km above the ground surface.
Actually, the constitution ratio of these photons varies according to the distance between the cloud base and the ground, ranging from 85\% at a distance of 5 km to 95\% at a distance of 0.3 km~\cite{tsuchiya_observation_2012}. Therefore, the present bursts exhibit constitution ratios of 85\%$-$95\% if the cloud-base altitude is in between 0.3 and 5 km.

\subsection{Duration}
Figure~\ref{fig:duration} presents the duration distribution for the final burst set.
Most of the observed duration ranged between 10 and 40 minutes, occasionally extending to $\sim$60 minutes.
This trend agrees with the typical life cycle of thunderclouds, 30$-$60 min, during the presummer and summer seasons. 
Their mature stage, which is likely to produce lightning, generally persists
for 15 to 30 min~\citep{ferrierhouzeOnedim1989}. 

The ARGO-YBJ experiment, operating $\sim$200 m away from the NM location, observed  
variations of the air-shower rate during 2012~\cite{argo_2022}. These variations lasted 14
to 55 min, and are in good agreement with our results.
Count enhancements with similar duration have been identified on Mt. Aragats (3250 m a.s.l.) and Mt. Lomnický Štít (2634 m a.s.l.), using multiple detectors such as NMs, a NaI detector, and a plastic scintillation detector. The count increases obtained by NMs~\cite{chilingarian_groundbased_2010,chilingarian_particle_2011} lasted $\sim$ 10 min, and most enhancements of the NaI count in energy $>$ 3 MeV lasted for a few minutes to 50 min, with rare occasions extending to 120 min~\cite{chilingarian_catalog_2019}. Similarly, those obtained at 
Mt. Lomnický Štít lasted for 1 min to 15 min~\cite{chum_significant_2020}. The typical durations
within 50 min are in agreement with the present result. However, the $>$120 min duration
has never been observed at the Yangbajing observatory. This discrepancy in burst duration may be attributed to two factors. 
One is the different meteorological conditions between Mt.~Aragats (3250 m a.s.l) and Yangbajing (4300 m a.s.l.): for example, successive generations of cells in thunderclouds might be more pronounced on Mt. Aragats. 
This difference may result in different life cycles for thundercloud-charged layers, where electron acceleration takes place. 
The other is a difference in the energy threshold for photons of 7 MeV for the Yangbajing NM and 3 MeV for their NaI detector. This difference in energy thresholds could also lead to different duration behaviors.

In contrast to high-altitude observations made during the pre-summer or summer seasons, 
observations conducted at the Japan Sea in winter seasons have revealed an alternative perspective on long-duration bursts, which generally last only a few minutes~\cite{torii_observation_2002,tsuchiya_detection_2007,tsuchiya_long-duration_2011,wada_catalog_2021}.
According to~\citet{KM_1994}, tripole charge structures become apparent during the mature stages of winter thunderclouds. 
These tripole structures persist for less than 10 min during early or late winter
and for even shorter duration, less than a few minutes during midwinter.
Thus, it is reasonable to postulate that differences in the life cycles of thunderclouds are responsible for the noticeable difference in burst duration among long-duration bursts during the summer and winter seasons. 
\subsection{Diurnal variation}
The distribution of the onset times for the final burst set is shown in Figure~\ref{fig:start_time}. This distribution is clearly divided into two sections: one covering the period from 00:00 to 10:00 local time (LT) and 
the other covering 10:00 to 24:00 LT.
LT is calculated as 6h $+$ universal time (UT), considering the geographical longitude of $90^\circ$. 
In the total number of observed bursts, 11\% of them began at 00:00-10:00 LT and the rest
occurred at 10:00-24:00 LT. 
The latter time region contains the period of increased convective activity in convective clouds over the TP 
as reported by \citet{uyedaCharacConvec2001} which showed that strong updrafts generated by land surfaces heated due to solar radiation in TP result in the rapid development of convective clouds during 09:00$-$13:00 LT 
in the premonsoon seasons and 12:00$-$16:00 LT in the monsoon seasons. In addition, \citet{qie_lightning_2003} investigated lightning activity between 1998 and 2002 in TP using the lightning image sensor onboard the Tropical Rainfall Measuring Mission. Their investigation indicates that
6\% and 78\% of lightning flashes in the diurnal lightning activity occur in 00:00$-$10:00 LT and 12:00$-$19:00 LT, respectively. Therefore, a clear correlation is evident between the distribution of the observed burst start times 
and the temporal pattern of lightning flashes.

\citet{wada_catalog_2021} examined the onset time distribution of 70 long-duration bursts detected in the coastal area of the Japan Sea over 4 winter seasons from 2015 to 2018. 
They divided the data into two parts: the daytime spanning 06:00$-$18:00 LT and the nighttime encompassing the remaining time period.
They showed that the number of events in  daytime is 15 and significantly smaller than 
50, the number of events in nighttime.
The present study also shows a comparable trend, indicating that the number of events during the day, 43, is less than the number of events during the night, 84, with a statistical significance of 3.6$\sigma$. 
This trend is supported to some extent by observations at Mt. Aragats~\cite{chilingarian_catalog_2019}.
These results suggest that long-duration bursts of high energy radiation are likely to occur during local nighttime hours. 
The nighttime phase corresponds to the dissipation phase of thunderclouds, 
during which the occurrence of strong electric fields capable of causing relativistic runaway electron avalanches is less likely. Therefore, 
it is plausible that the long-duration bursts observed at night could be attributed to the process of Modification Of Spectra as discussed by \citet{Chilingarian_MOS_2014} and \citet{dinizatomoelejgr2022}.

\subsection{Seasonal variation}
Figure~\ref{fig:mth_var} shows the seasonal variation in the occurrences number of the final burst set. 
Among them, 122 events (97\% of the total 127 events) were observed between May and September, corresponding to the premonsoon and monsoon seasons
in TP. The seasonal trend obtained is similar to
that of precipitation and lightning activity within the TP region~\cite{qie_lightning_2003}. This similarity suggests
that long-duration bursts tend to occur during the rainy season when lightning is more frequent, as expected. 

According to the EF measurements from March 2010 to February 2012~\cite{amenomori_observation_2013}, the total number of thundercloud occurrences at Yangbajing was 158 days. During this period, the Yangbajing NM detected only seven long-duration bursts (Table~\ref{tab:eventslist}). Interestingly, \citet{amenomori_observation_2013} showed that thunderclouds were also observed in 54 days during winter seasons from December to March. However, the Yabgbajing NM did not
detect any long-duration bursts during this winter period, except only one recorded 
between 1998 and 2017 (20040203 in Table~\ref{tab:eventslist}).
This infrequent detection during winter suggests that the EF intensity due to thunderclouds may not be the sole factor governing the production of long-duration bursts.

The ARGO-YBJ experiment, conducted near the NM location, detected twenty count enhancements during thunderstorms in 2012 between April and October~\cite{argo_2022}. This period coincides with the present result. Notably, four (20120714, 20120804, 20120911, and 20121010) of the six long-duration bursts detected in 2012 (Table~\ref{tab:eventslist}) align with the dates recorded by the ARGO-YBJ experiment. 
The effective area of the Yangbajing NM (32 $\mathrm{m^2}$) is smaller than that of the ARGO-YBJ 
detector ($\sim$$110\times100$ $\mathrm{m^2}$). A larger effective area is likely to detect more long-duration bursts with narrow illumination areas. Hence the difference in effective area is considered one of the main factors contributing to the variation in the number of events detected in 2012 (6 for the Yangbajing NM and 20 for the ARGO-YBJ experiment).

\citet{muraki_effects_2004} studied a seasonal variation using a large-area meson monitor installed at Mt.~Norikura (2770 m a.s.l.) in Japan. The meson monitor comprises scintillation detectors, and
primarily detects charged particles but also has some sensitivity to neutral particles such as neutrons and gamma rays. 
Analyzing data obtained from October 1990 to January 2002, they found that the derived seasonal variation matches the pattern of thunderstorm occurrence. The seasonal variation observed with the Yangbajing NM is similar to their result. This is probably because the two areas are influenced by the Asian monsoon. 
\citet{chilingarian_catalog_2019} also showed a seasonal variation, with a notable peak in May, which may probably be due to different meteorological conditions compared to those observed in the TP and at Mt.~Norikura.

\subsection{Annual variation}
In Figure~\ref{fig:annual_ev}, we present an annual variation in occurrences numbers of the final 
burst set (red histogram in the top panel)
and compare it with the annual variations of the sunspot number (black histogram in the top panel) 
and the galactic cosmic-ray intensity (histogram in the bottom panel) to explore their possible 
correlations with long-duration bursts. The sunspot number data were sourced from Sunspot Index and Long-term Solar Observations: $\mathrm{https://www.sidc.be/silso/home}$. For our investigation, we used Jungfraujoch NM data that are available from the World Data Centre for Cosmic Rays ($\mathrm{https://cidas.isee.nagoya-u.ac.jp/WDCCR}$), but annual variations of galactic cosmic ray intensity are almost common in all NM data.
It has been hypothesized that galactic cosmic rays play a role in the induction of thunderstorm-related events by providing seed electrons~\citep{gurevich_runaway_1992,babich_study_2007,carlson_runaway_2008,gurevich_runaway_2013}. As is well known, variations in sunspot number reflect solar activity that has the 11-year cycle, and is inversely correlated with variations of galactic cosmic ray intensity at Earth.
Thus, it is natural to conjecture that solar activity may influence the occurrence of long-duration bursts. 

To quantify the annual variation of the occurrence rate in the final burst set, 
we employed the following functions for fitting: (1) a constant function, (2) a linear function, and (3) a periodic function. 
Functions (1) and (2) are denoted $f_c(t) = a_0$ and $f_l(t) = a_0 + a_1 t$, respectively. 
Here $a_0$ and $a_1$ represent a constant and a slope, respectively. The variable $t$ shows the
elapsed year since 1998.
The function (1) assumes that solar activity or cosmic rays exert time-independent effects on the observed number of long-duration bursts.
Function (2) assumes a descending (or ascending) trend during the 1998$-$2017 period, based on the fact that
the peak sunspot number was smaller around 2013$-$2015 than during the period around 2000$-$2002.
The last periodic function (3) is expressed as 
\begin{equation}
f_\mathrm{p}(t) =  a_{0} + A\sin\left(\frac{2\pi}{T}t+\phi \right), 
\label{eq:p1}
\end{equation}
where $A$, $T$ and $\phi$ denote a normalization constant, a period, and a phase, respectively.
As listed in Table~\ref{tab:yr_fit_res}, the resultant $\chi^2/\nu$ values ($\nu$ is the degree of freedom) 
indicate that $f_p(t)$ provides the best fit to reproduce the derived annual variation of
the final burst set. This result implies the potential presence of periodicity in the annual occurrence rate
of the observed long-duration bursts.

For comparison, we evaluated periods of annual variations in the sunspot number and cosmic ray flux as well
by fitting the following function to the variations
\begin{equation}
f_\mathrm{sun,cr}(t) =  a_{0} + a_{1}t + A\sin\left(\frac{2\pi}{T}t+\phi \right). 
\label{eq:p2}
\end{equation}
This function considers a clear decrease (increase) in the amplitude of the sunspot number variation 
(cosmic-ray flux) over the years. The resulting curves of $f_{\mathrm{sun}}(t)$ and $f_\mathrm{cr}(t)$ are drawn in 
Figure~\ref{fig:annual_ev}, and the parameters determined are listed in Table~\ref{tab:yr_fit_res}. 
The evaluated periods ($T$) for the sunspot number and cosmic-ray flux
deviate from the period estimated from the final burst set by 2.2$\sigma$ and 2.7$\sigma$ (Table~\ref{tab:yr_fit_res}).
In addition, by using the derived periods ($T$) and phases ($\phi$) in Table~\ref{tab:yr_fit_res},
we can calculate the peak years of $f_\mathrm{p}$ and $f_\mathrm{sun}$, and the year of the minimum of $f_\mathrm{cr}$,
each measured from 1998, as $6.9\pm1.3$ years, $3.58\pm0.13$ years, and $4.756 \pm 0.003$ years, respectively. From these values, possible slight
lags of $f_\mathrm{p}$ with respect to $f_\mathrm{sun}$ and  
$f_\mathrm{cr}$ were estimated as $3.3\pm1.3$ years and $2.1\pm1.3$ years, respectively.
While the present analysis suggests a potential connection between long-duration bursts and solar activity or cosmic rays, 
we need more data over a longer time period before definitively confirming the periodicity and delay.


\section{Conclusions}
Long-duration bursts were detected by the Yangbajing NM located at 4300 m a.s.l.
during the period of 1998$-$2017. The detected bursts have a typical duration of 10 to 40 min. 
This duration distribution suggests that electrons undergo continuous acceleration to 10 MeV or higher within 
a few tens of minutes. In the coastal area of the Japan Sea, long-duration bursts have also been observed in winter
seasons and typically last for a few minutes. The difference in duration
would result from the distinct life cycles of winter and summer thunderclouds.

The seasonal and diurnal variations in the occurrence number of long-duration bursts 
agreed well with those variations of lightning and precipitation on the TP.
This agreement justifies the selection and subsequent analysis of the Yangbajing NM data. 
The present work, based on 19-year data, suggests the possibility that 
the annual variation of long-duration 
bursts has a periodicity of $\sim$16 years and a lag of $\sim$3 years relative to solar activity.
To gain further insight into the observed annual variability, 
it would be necessary to conduct longer-term high-energy radiation observations coupled with comprehensive meteorological measurements.

The present work highlights the importance of NM data for understanding the temporal variability and
underlying production mechanisms of long-duration bursts of high energy radiation associated with thunderclouds.
Since the 1950s, numerous NMs have been operational at various observatories around the world~\cite{butikoferGroundBasedMeasurementsEnergetic2018a}. 
These NM data would allow us to unravel the real connections between long-duration bursts, cosmic rays, and solar activity.


\begin{backmatter}

\section*{Abbreviations}
TP: Tibetan plateau; NM: Neutron monitor; EFM: Electric filed mill; 

\section*{Availability of data and material}


The datasets used and/or analysed during the current study are available from the corresponding author on reasonable request.






\section*{Competing interests}

The authors declare that they have no competing interests.

\section*{Funding}

This work was partly supported by JSPS KAKENHI Grant Number 21H01116.

\section*{Authors' contributions}


HT, KH, and KK were responsible for data analysis, interpretation, and the proposed writing of this paper. 
KH and KK installed the EFMs and performed their maintenance. HT, KM, CK, MO, MT, SS QS, SW, CH, and LZ contributed to the maintenance of the Yangbajing NM.  All authors read and approved the final manuscript.





\section*{Acknowledgements}
We sincerely thank Prof. T. Shinoda (Nagoya Univ.) for giving information on 
meteorological conditions on the Tibetan plateau. 
\bibliographystyle{plainnat} 
\bibliography{ref20231020}  

 %
%
%

\newpage





%
\begin{figure}[h!]
\includegraphics[scale=0.20]{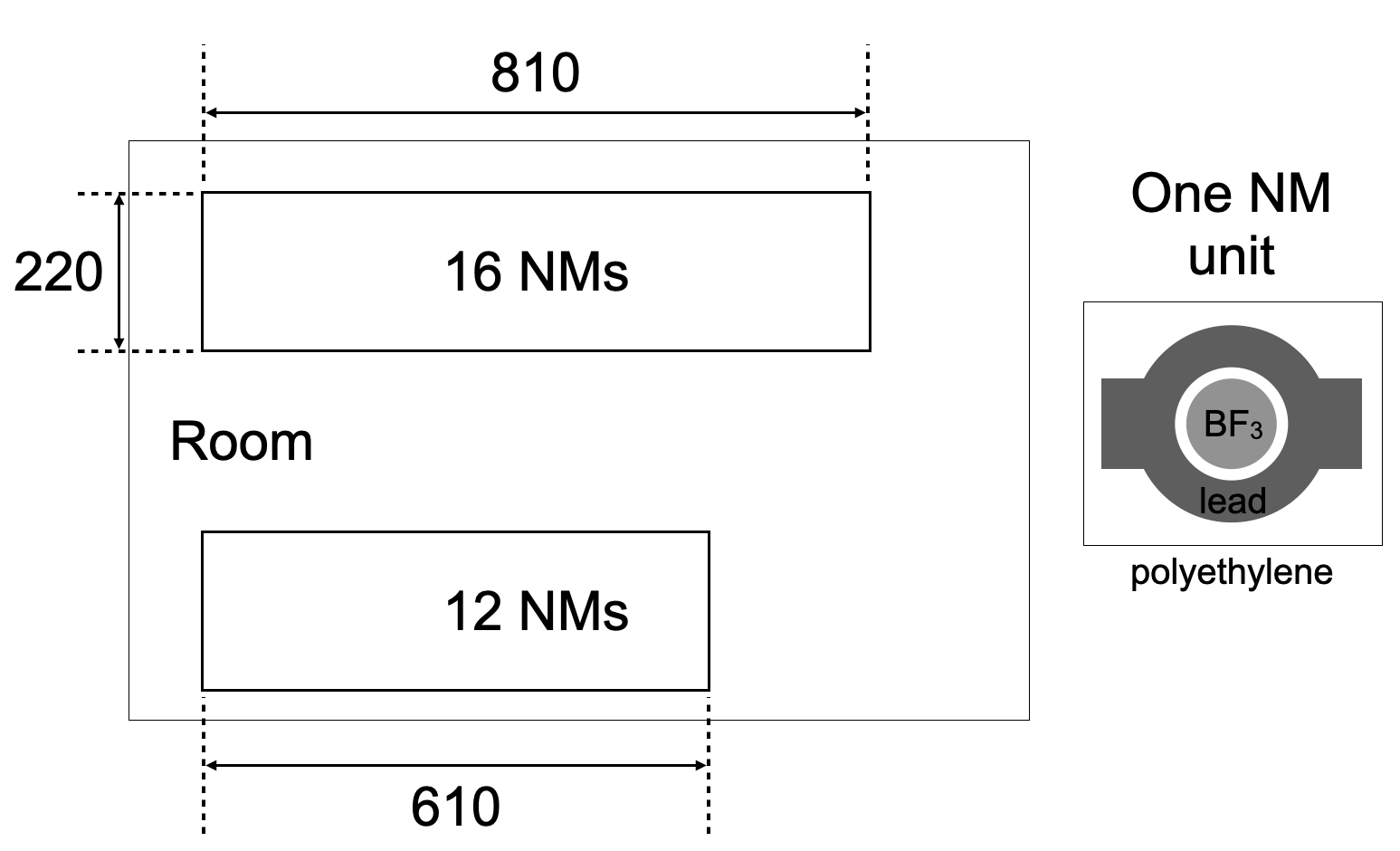}
\caption{Arrangement of the Yangbajing NM in a room and a cross-sectional view of one unit, measured in centimeters.}
\label{fig:NMscmatic}
\end{figure}
\clearpage

\begin{figure}[h!]
\includegraphics[scale=0.15]{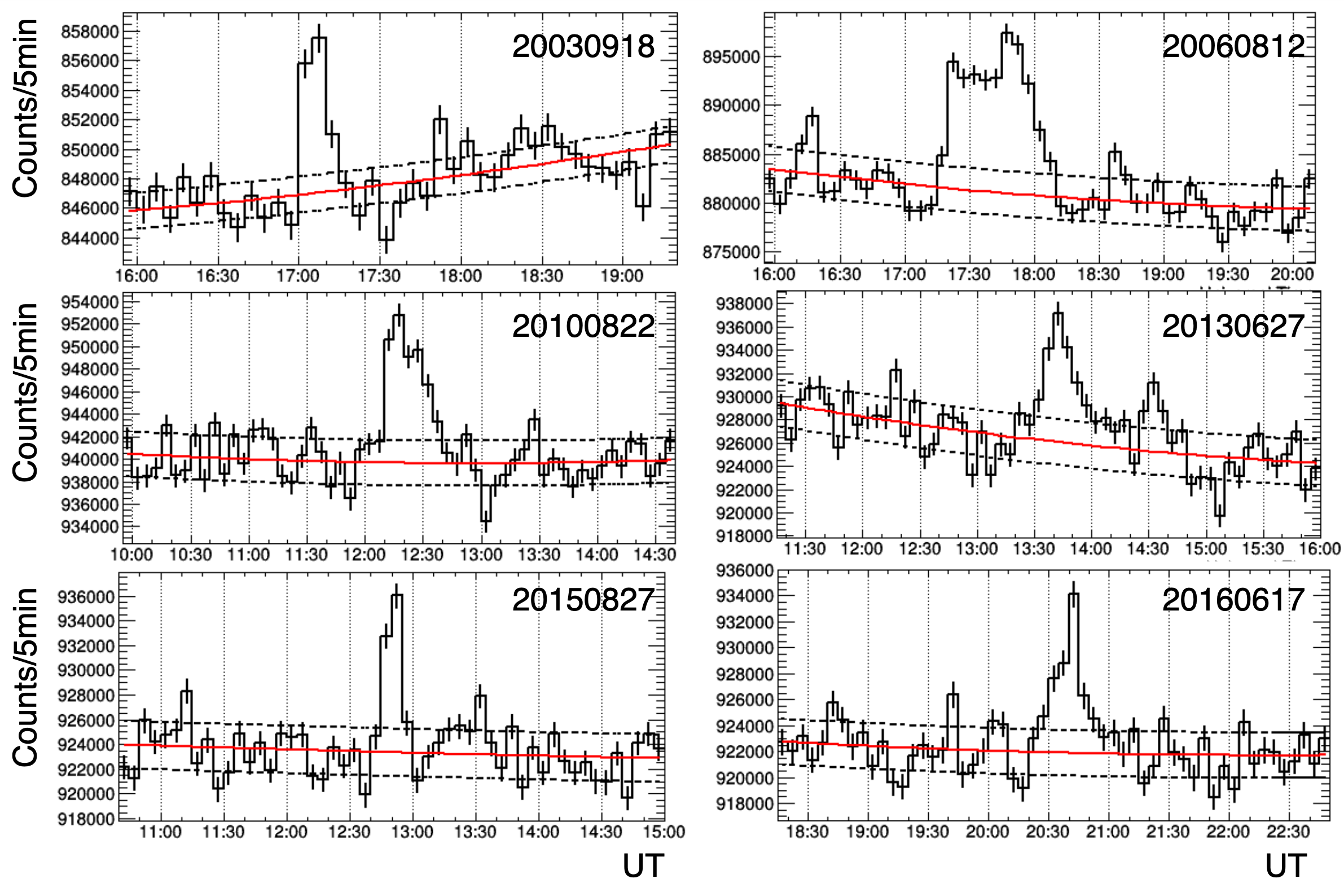}
 \caption{Examples of the NM count rate histories per 5 minutes in the final burst set.
Each panel is labeled with the date of the event acquisition. 
The horizontal axis depicts universal time (UT). Each error shown corresponds to a statistical 1$\sigma$.
Red solid lines represent estimated background levels, while two black dashed lines indicate
a range of $\pm \sigma_\mathrm{1day}$ from the background level.}
\label{fig:exampleLCs}
\end{figure}
\clearpage
%
%
\begin{figure}[h!]
\includegraphics[scale=0.30]{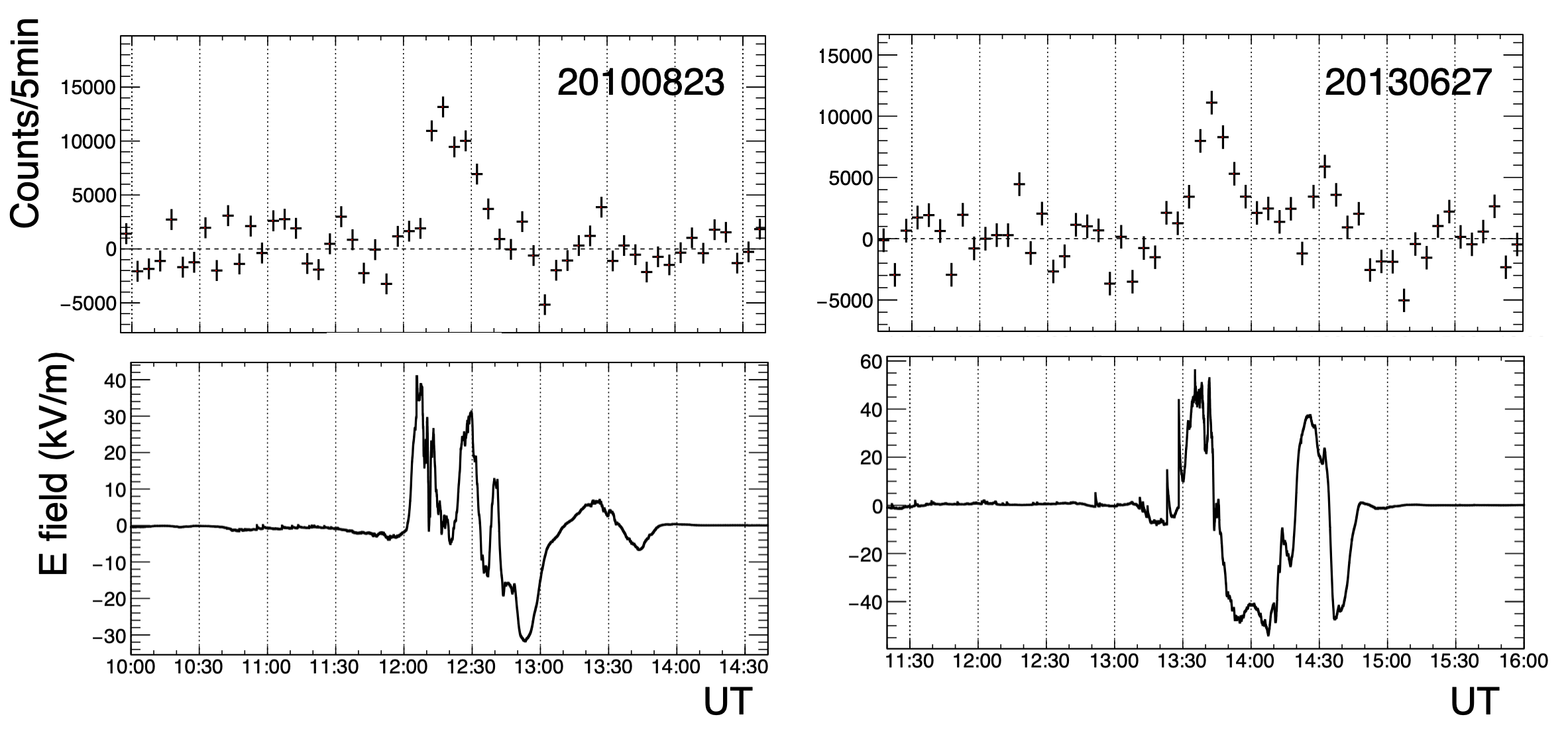}
\includegraphics[scale=0.22]{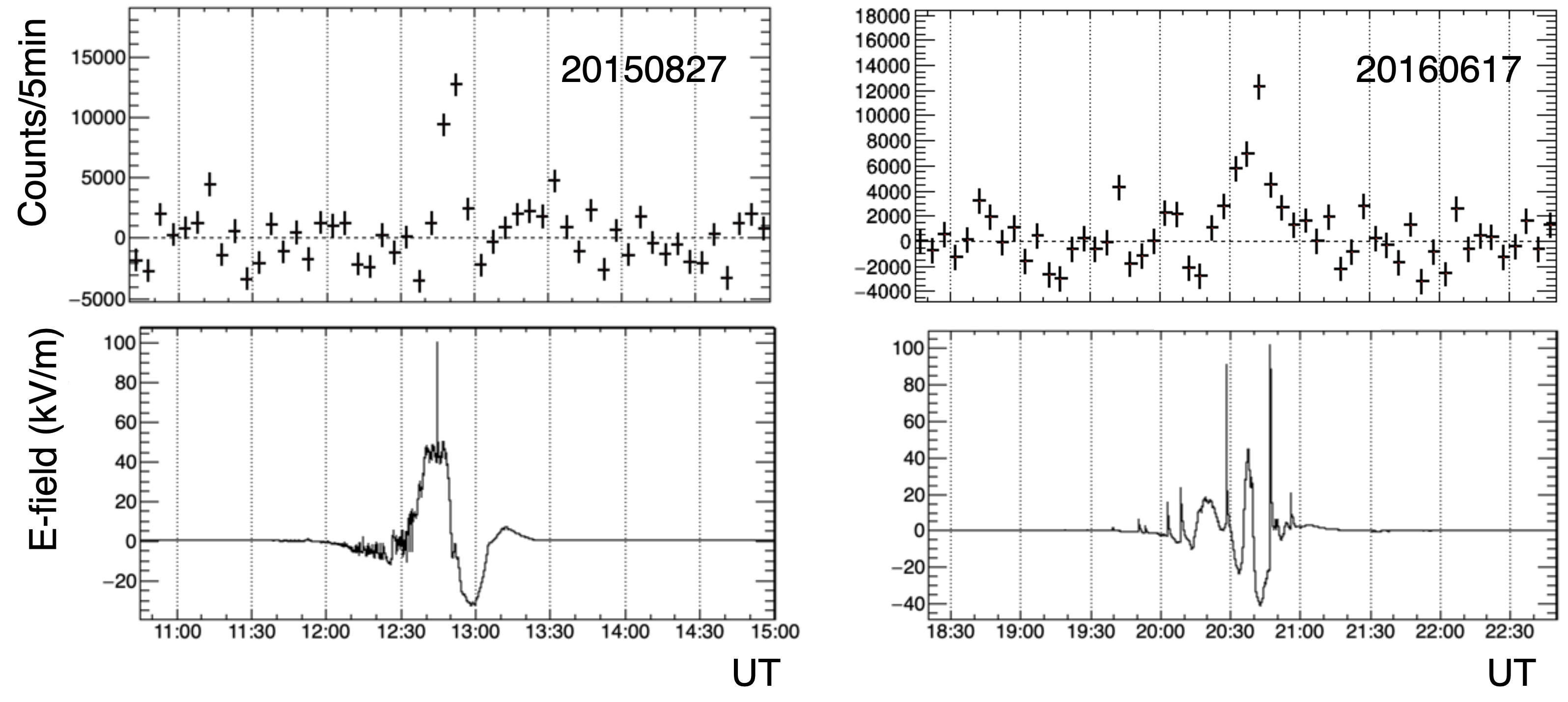}
 \caption{Comparison between the BG-subtracted NM count rate and electric field (solid line). 
The abscissa depicts universal time (UT). The errors quoted from the NM counts are statistical 1$\sigma$.
}
\label{fig:TwoCntWithEFMex1}
\end{figure}
\clearpage

%
%
%
%

%
%
\begin{figure}[h!]
\includegraphics[scale=0.5]{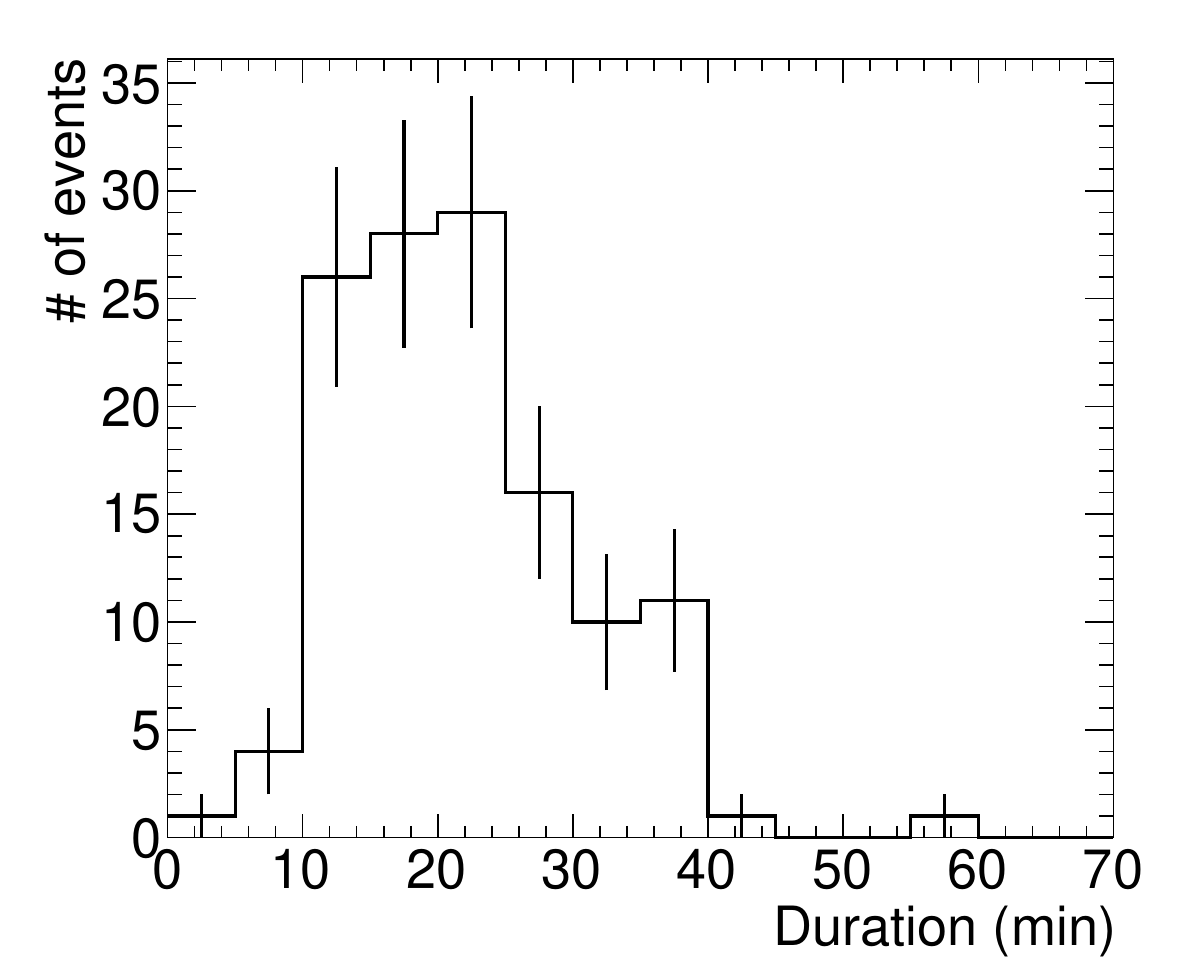}
\caption{Duration distribution of long-duration bursts in the final burst set.}
\label{fig:duration}
\end{figure}
%
%
%
\begin{figure}[h!]
\includegraphics[scale=0.50]{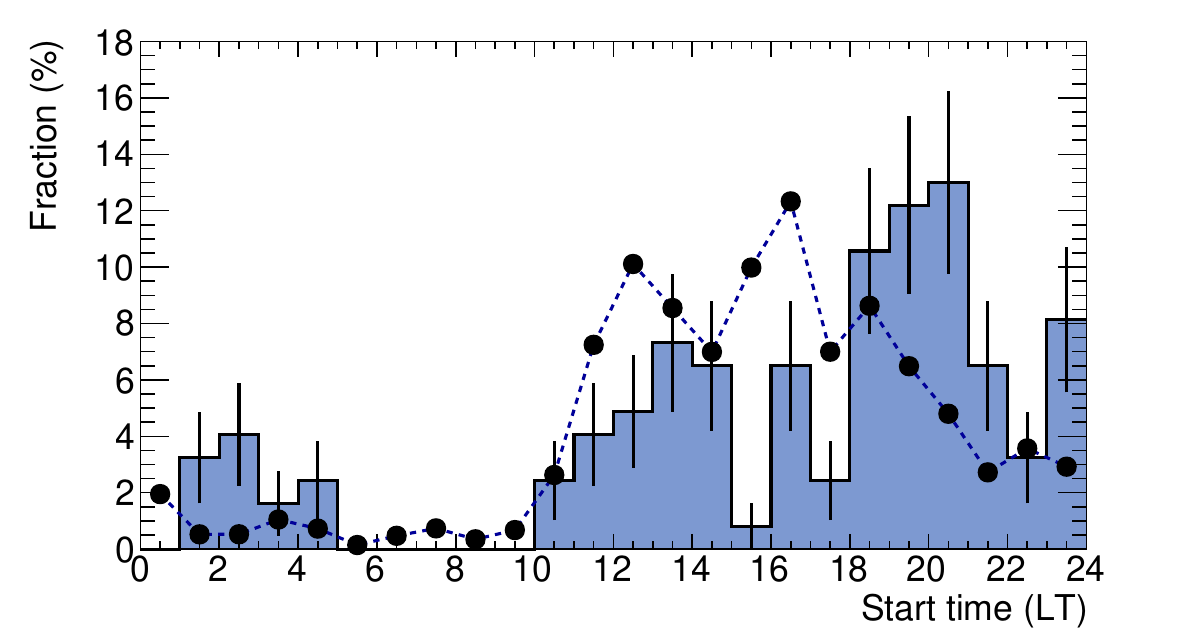}
\caption{Start time distribution of events in the final burst set (Filled histogram), plotted as a function of local time (LT) computed by 6h + UT. The filled circles represent the diurnal variation of
lightning activity in TP in 1998$-$2002~\cite{qie_lightning_2003}.}
\label{fig:start_time}
\end{figure}
%
\begin{figure}[h!]
\includegraphics[scale=0.5]{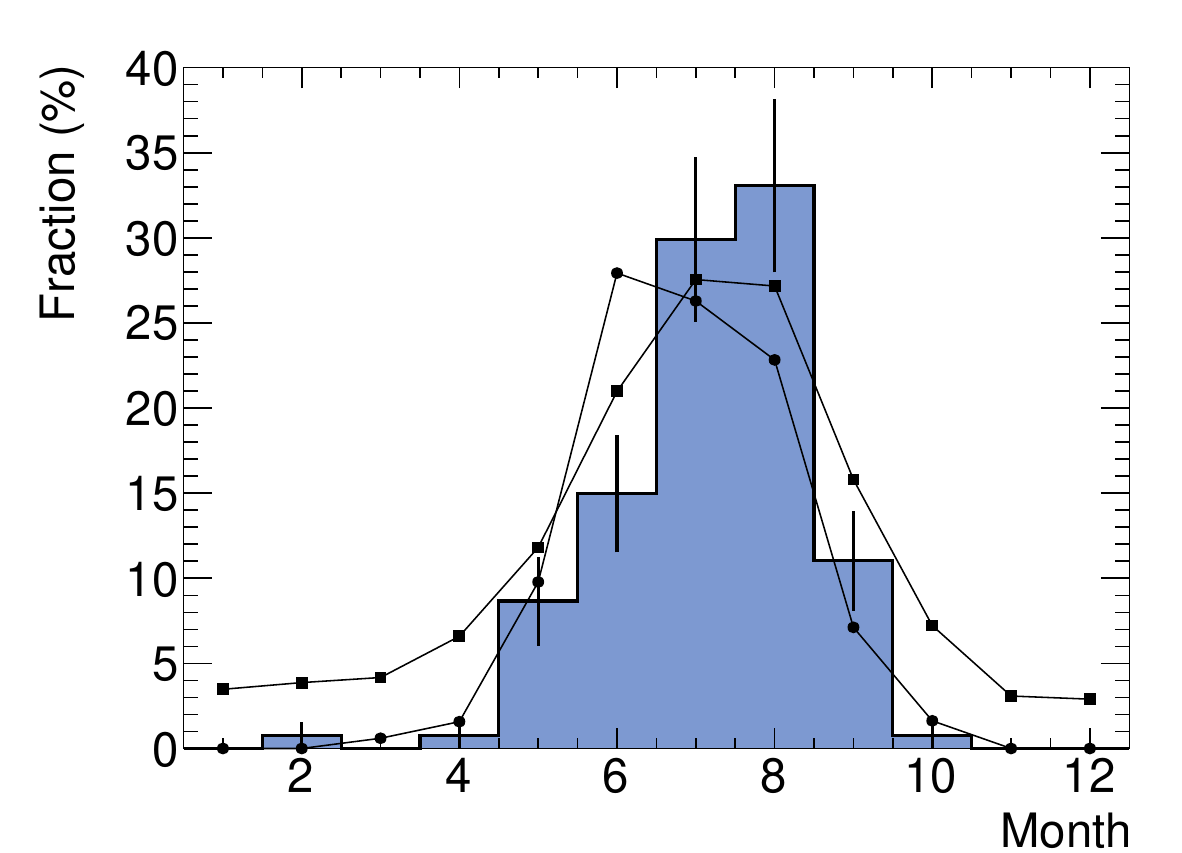}
 \caption{Distribution of months in which events occurred in the final burst set (filled histogram). Filled squares and circles indicate lightning activity and precipitation, respectively, during the years 1998$-$2002 
 in the central part of the TP~\citep{qie_lightning_2003}.}
\label{fig:mth_var}
\end{figure}
\clearpage
%
%
\begin{figure}[h!]
\includegraphics[scale=0.4]{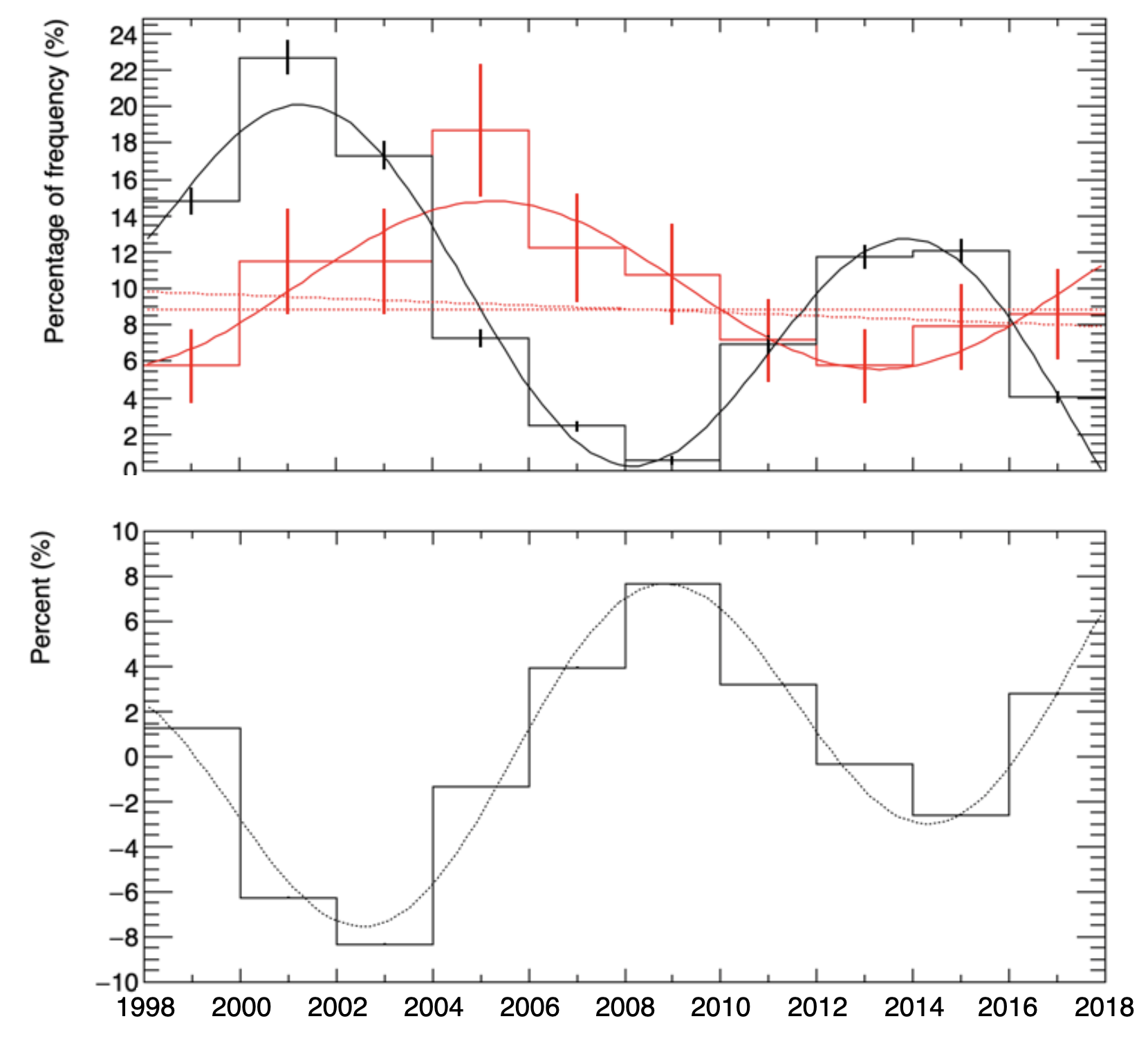}
\caption{Top: annual occurrence variation of the final data set (red histogram). The black histogram shows an annual variation in sunspot number expressed in \%.
Solid red and black curves denote a periodic function of $f_p(t)$ and $f_{sun}(t)$, respectively, while dashed lines represent $f_c(t)$ and $f_l(t)$.The error bar represents statistical 1$\sigma$.Bottom: the annual variation of cosmic-ray flux measured by neutron monitors located at Jungfraujoch. The NM data were
obtained from the following URL:https://cidas.isee.nagoya-u.ac.jp/WDCCR/index.html.
}
\label{fig:annual_ev}
\end{figure}

%
%
%

\clearpage
\begin{longtable}[c]{cccccc}
\caption{The parameters determined for individual events in the final burst set.}
\label{tab:eventslist}
\\
\hline
date (YYYYMMDD)   & $t_\mathrm{s}$ (HHMM)\tablefootnote{start time}  & $t_\mathrm{e}$(HHMM)\tablefootnote{end time} & $\Delta t_\mathrm{d}$  & $N_\mathrm{tot}$ & significance\tablefootnote{significance of $N_\mathrm{tot}$}  \\
         &     (UT)               &   (UT)               &        (s)             &                  &   \\
\hline\hline
\endfirsthead
\hline
\multicolumn{6}{c}{\small\it continued}\\
date (YYYYMMDD)  & $t_\mathrm{s}$ (HHMM)  & $t_\mathrm{e}$(HHMM) & $\Delta t_\mathrm{d}$  & $N_\mathrm{tot}$ & significance \\
         &     (UT)               &   (UT)               &        (s)             &                  &   \\
\hline\hline
\endhead
%
\hline
\multicolumn{6}{c}{\small\it next page}\\
\endfoot
\hline
\multicolumn{6}{c}{\small\it end }\\
\endlastfoot
 
19990526 & 1130 & 1134 &  240 &  24866 & 15.8 \\ 
19990609 & 1430 & 1455 & 1500 &  35669 &  7.7 \\ 
19990620 & 0730 & 0755 & 1500 &  24092 &  6.9 \\ 
19990629 & 1910 & 1925 &  900 &  17985 &  7.9 \\ 
19990808 & 1405 & 1425 & 1200 &  15918 &  5.1 \\ 
19990818 & 1330 & 1405 & 2100 &  32409 &  6.9 \\ 
19990830 & 1300 & 1340 & 2400 &  41992 &  9.2 \\ 
19990905 & 1355 & 1425 & 1800 &  40935 & 12.0 \\ 
20000523 & 1950 & 2025 & 2100 &  39599 & 10.7 \\ 
20000527 & 1040 & 1050 &  600 &  13721 &  6.5 \\ 
20000701 & 1820 & 1835 &  900 &  31168 & 10.8 \\ 
20000713 & 0835 & 0900 & 1500 &  44507 &  8.4 \\ 
20000720 & 1225 & 1245 & 1200 &  25997 &  8.0 \\ 
20000729 & 1450 & 1515 & 1500 &  26050 &  6.7 \\ 
20000816 & 0820 & 0855 & 2100 &  39777 &  6.9 \\ 
20000818 & 1320 & 1330 &  600 &  22382 &  6.4 \\ 
20000818 & 1400 & 1415 &  900 &  14774 &  5.2 \\ 
20000822 & 0800 & 0820 & 1200 &  28867 &  9.4 \\ 
20010426 & 0715 & 0730 &  900 &  19924 &  8.8 \\ 
20010609 & 1700 & 1735 & 2100 &  51037 & 11.4 \\ 
20010620 & 1325 & 1345 & 1200 &  27694 &  7.2 \\ 
20010625 & 1205 & 1215 &  600 &  11387 &  5.8 \\ 
20010801 & 2005 & 2020 &  900 &  26538 & 10.0 \\ 
20020518 & 1410 & 1420 &  600 &  12270 &  4.4 \\ 
20020519 & 1650 & 1710 & 1200 &  14790 &  5.0 \\ 
20020529 & 1550 & 1600 &  600 &  13043 &  6.0 \\ 
20020602 & 2105 & 2115 &  600 &  15976 &  7.3 \\ 
20020730 & 1610 & 1630 & 1200 &  31907 &  7.8 \\ 
20030620 & 1434 & 1440 &  360 &  12416 &  4.1 \\ 
20030628 & 0950 & 1015 & 1500 &  26251 &  5.8 \\ 
20030630 & 1025 & 1050 & 1500 &  32280 &  7.3 \\ 
20030709 & 0540 & 0550 &  600 &  15861 &  6.3 \\ 
20030714 & 1530 & 1545 &  900 &  10919 &  4.0 \\ 
20030820 & 1455 & 1515 & 1200 &  17700 &  5.6 \\ 
20030822 & 1340 & 1405 & 1500 &  19572 &  6.3 \\ 
20030822 & 1430 & 1445 &  900 &  37628 &  9.4 \\ 
20030825 & 1250 & 1325 & 2100 &  46328 & 11.0 \\ 
20030918 & 1700 & 1715 &  900 &  23264 & 11.0 \\ 
20040203 & 1545 & 1615 & 1800 &  29535 &  8.0 \\ 
20040609 & 2240 & 2250 &  600 &  26507 & 11.1 \\ 
20040704 & 2025 & 2040 &  900 &  22875 &  6.6 \\ 
20040725 & 0735 & 0800 & 1500 &  28833 &  8.6 \\ 
20040731 & 1425 & 1445 & 1200 &  20889 &  4.7 \\ 
20040803 & 1625 & 1645 & 1200 &  22040 &  5.4 \\ 
20040804 & 2225 & 2255 & 1800 &  38885 &  9.0 \\ 
20040809 & 1325 & 1335 &  600 &  18530 & 10.3 \\ 
20040811 & 0825 & 0845 & 1200 &  44852 & 11.4 \\ 
20040811 & 1725 & 1750 & 1500 &  26480 &  7.5 \\ 
20040830 & 1600 & 1615 &  900 &  13001 &  5.2 \\ 
20040901 & 1535 & 1555 & 1200 &  12259 &  4.0 \\ 
20040904 & 1215 & 1235 & 1200 &  22654 &  6.6 \\ 
20050508 & 1210 & 1230 & 1200 &  16474 &  5.9 \\ 
20050605 & 0450 & 0510 & 1200 &  23276 &  7.6 \\ 
20050615 & 1320 & 1340 & 1200 &  20656 &  5.5 \\ 
20050702 & 1225 & 1235 &  600 &  47259 &  6.0 \\ 
20050724 & 1520 & 1545 & 1500 &   9991 &  4.2 \\ 
20050725 & 1835 & 1900 & 1500 &  29619 &  6.4 \\ 
20050807 & 1410 & 1425 &  900 &  48742 & 13.2 \\ 
20050807 & 1430 & 1445 &  900 &  13820 &  4.3 \\ 
20050826 & 0720 & 0740 & 1200 &  16826 &  5.6 \\ 
20050903 & 1120 & 1140 & 1200 &  22081 &  7.5 \\ 
20050917 & 0710 & 0720 &  600 &   9444 &  4.3 \\ 
20060511 & 1748 & 1754 &  360 &  10765 &  5.7 \\ 
20060528 & 0535 & 0545 &  600 &  15920 &  7.0 \\ 
20060627 & 1235 & 1250 &  900 &  15087 &  4.1 \\ 
20060628 & 0555 & 0610 &  900 &  18245 &  5.8 \\ 
20060808 & 1730 & 1740 &  600 &  23705 & 11.0 \\ 
20060812 & 1715 & 1810 & 3300 & 113029 & 15.7 \\ 
20060829 & 0730 & 0800 & 1800 &  57789 & 15.0 \\ 
20060922 & 0615 & 0625 &  600 &  17815 &  6.0 \\ 
20060924 & 0839 & 0845 &  360 &   9566 &  5.4 \\ 
20070701 & 1840 & 1900 & 1200 &  30533 &  9.8 \\ 
20070713 & 1430 & 1445 &  900 &  22101 &  7.1 \\ 
20070722 & 0615 & 0630 &  900 &  22137 &  7.8 \\ 
20070727 & 1520 & 1540 & 1200 &  20633 &  5.6 \\ 
20070824 & 1150 & 1200 &  600 &  16041 &  5.5 \\ 
20070825 & 1455 & 1530 & 2100 &  86911 & 15.4 \\ 
20080515 & 1525 & 1545 & 1200 &  20732 &  6.4 \\ 
20080531 & 2125 & 2150 & 1500 &  33840 & 10.7 \\ 
20080621 & 1945 & 1955 &  600 &  18923 &  7.1 \\ 
20080711 & 0640 & 0650 &  600 &  14146 &  4.8 \\ 
20080714 & 1225 & 1240 &  900 &  19914 &  5.9 \\ 
20080827 & 1000 & 1030 & 1800 &  42310 &  8.5 \\ 
20090628 & 1230 & 1240 &  600 &  13239 &  5.1 \\ 
20090702 & 1000 & 1015 &  900 &   9734 &  4.0 \\ 
20090702 & 1020 & 1030 &  600 &  19118 &  6.4 \\ 
20090719 & 2200 & 2220 & 1200 &  43259 & 13.5 \\ 
20090725 & 0650 & 0715 & 1500 &  21689 &  5.8 \\ 
20090801 & 0540 & 0600 & 1200 &  15542 &  4.5 \\ 
20090802 & 1355 & 1420 & 1500 &  26839 &  7.5 \\ 
20090906 & 1000 & 1025 & 1500 &  25122 &  5.8 \\ 
20100706 & 1855 & 1910 &  900 &  15802 &  5.8 \\ 
20100722 & 0445 & 0505 & 1200 &  37639 & 12.0 \\ 
20100823 & 1210 & 1240 & 1800 &  54236 & 11.1 \\ 
20100919 & 1420 & 1440 & 1200 &  30433 &  8.6 \\ 
20100924 & 0750 & 0820 & 1800 &  45447 & 11.8 \\ 
20110710 & 1310 & 1345 & 2100 &  29362 &  8.3 \\ 
20110811 & 1735 & 1745 &  600 &  26710 &  8.2 \\ 
20120711 & 0605 & 0625 & 1200 &  30299 &  8.5 \\ 
20120714 & 1750 & 1805 &  900 &  21427 &  6.2 \\ 
20120803 & 2010 & 2025 &  900 &  33062 &  9.6 \\ 
20120804 & 1730 & 1750 & 1200 &  23692 &  7.3 \\ 
20120911 & 0550 & 0615 & 1500 &  25331 &  6.5 \\ 
20121010 & 0645 & 0655 &  600 &  19303 &  6.0 \\ 
20130627 & 1330 & 1400 & 1800 &  44102 &  7.9 \\ 
20130902 & 0810 & 0825 &  900 &  20738 &  8.0 \\ 
20140703 & 1025 & 1040 &  900 &  30253 &  5.8 \\ 
20140708 & 0755 & 0805 &  600 &  15058 &  5.4 \\ 
20140716 & 0835 & 0850 &  900 &  16548 &  5.4 \\ 
20140720 & 1355 & 1405 &  600 &  31341 &  4.8 \\ 
20140810 & 0405 & 0425 & 1200 &  27330 &  5.3 \\ 
20140810 & 1900 & 1915 &  900 &  41085 &  9.8 \\ 
20140821 & 1516 & 1522 &  360 &  12142 &  4.5 \\ 
20140823 & 1040 & 1055 &  900 &  30624 &  6.5 \\ 
20140911 & 1435 & 1455 & 1200 &  37654 & 10.2 \\ 
20150827 & 1245 & 1300 &  900 &  19220 &  5.7 \\ 
20160617 & 2025 & 2055 & 1800 &  24536 &  7.4 \\ 
20160702 & 1755 & 1810 &  900 &  32135 &  8.2 \\ 
20160704 & 1435 & 1510 & 2100 &  18502 &  6.3 \\ 
20160710 & 1210 & 1220 &  600 &  41234 &  7.5 \\ 
20160717 & 1220 & 1255 & 2100 &  24763 &  5.8 \\ 
20160717 & 1348 & 1400 &  720 &  39823 &  8.5 \\ 
20160801 & 0700 & 0720 & 1200 &  29004 &  6.5 \\ 
20170705 & 1330 & 1340 &  600 &  17946 &  6.3 \\ 
20170816 & 1305 & 1340 & 2100 &  50353 & 15.4 \\ 
20170816 & 2030 & 2105 & 2100 &  46590 & 13.2 \\ 
20170820 & 0805 & 0835 & 1800 &  58489 & 11.7 \\ 
\end{longtable}
%
%
\begin{table}[ht!]
\caption{\label{tab:yr_fit_res}The fitting parameters determined for each function.}
\begin{tabular}{ccccccc}
         &    $A$   & $T$         & $\phi$       &  $a_0$      & $a_1$   &    $\chi^2/\nu$\tablefootnote{$\nu$ is degree of freedom.} \\      
Function &    ($\%$)   & (yr)    & (rad) &  (yr)  &  ($\times10^{-1}$\%/yr) & \\ 
$f_c(t)$ &    $-$   &$-$     & $-$    & $(8.8 \pm 0.8)\times 10^{-2}$  & $-$ & $16/9$ \\
$f_l(t)  $ &    $-$   &$-$     & $-$    &  $(1.0 \pm 0.2)\times 10^{-1}$ & $-1.5\pm 1.4$ & $14/8$\\             
$f_\mathrm{p}(t)  $ & $4.6\pm 1.3$ & $16.4 \pm 1.7$   & $-1.1 \pm 0.4$ & $0.100 \pm 0.009$ & $-$ & $3.9/6$ \\
$f_\mathrm{sun}(t)$ & $8.0\pm 0.2$ & $12.57 \pm 0.13$ & $-0.22 \pm  0.06$ & $0.141 \pm 0.004$& $-5.8 \pm 0.3$ & $33/5$ \\
$f_\mathrm{cr}(t)$ & $6.422\pm 0.009$ & $11.772 \pm 0.005$ & $2.174 \pm  0.002$ & $-2.920 \pm 0.013$& $ 0.388\pm 0.001$ & $50/5$ \\
\end{tabular}
\end{table}


\section*{Additional Files}
  \subsection*{Additional file 1}
This file includes comparisons between long-duration bursts detected in 2010-2017 (Table 1 in the main text) and EF variations, but not include the comparisons presented in the main text. Due to periods of the EFM shutdown, EF variations could not be obtained for all periods of 2010-2017.
%
%
\end{backmatter}
\end{document}